\documentclass[twocolumn,showpacs,english,prl,superscriptaddress,preprintnumbers,amsmath,amssymb,floatfix]{revtex4}

\usepackage[T1]{fontenc}
\usepackage[latin9]{inputenc}
\usepackage{epsfig,psfrag}
\usepackage{dcolumn}
\usepackage{bm}
\usepackage{graphicx}
\usepackage{color}
\usepackage{esint}
\usepackage{babel}
\usepackage{amsfonts}
\usepackage{enumerate}

\newcommand{\mean}[1]{\langle #1 \rangle}

\begin{document}

\title{Majorana single-charge transistor}

\author{R. H\"utzen}
\affiliation{ Institut f\"ur Theoretische Physik, 
Heinrich-Heine-Universit\"at, D-40225  D\"usseldorf, Germany }

\author{A. Zazunov}
\affiliation{ Institut f\"ur Theoretische Physik, 
Heinrich-Heine-Universit\"at, D-40225  D\"usseldorf, Germany }
 
\author{B. Braunecker}
\affiliation{Departamento de F{\'i}sica Te{\'o}rica de la 
Materia Condensada C-V and Instituto Nicol\'as Cabrera,
Universidad Aut{\'o}noma de Madrid, E-28049 Madrid, Spain }

\author{A. Levy Yeyati}
\affiliation{Departamento de F{\'i}sica Te{\'o}rica de la 
Materia Condensada C-V and Instituto Nicol\'as Cabrera,
Universidad Aut{\'o}noma de Madrid, E-28049 Madrid, Spain }

\author{R. Egger}
\affiliation{ Institut f\"ur Theoretische Physik, 
Heinrich-Heine-Universit\"at, D-40225  D\"usseldorf, Germany }

\date{\today}

\begin{abstract}
We study transport through a Coulomb blockaded topologically nontrivial 
superconducting wire (with Majorana end states) contacted by metallic leads.
An exact formula for the current through this interacting Majorana
single-charge transistor is derived in terms of wire spectral functions.
A comprehensive picture follows from three different approaches.
We find Coulomb oscillations with universal halving of the 
finite-temperature peak conductance under strong blockade conditions,
where the  valley conductance mainly comes from elastic cotunneling.
The nonlinear conductance exhibits finite-voltage sidebands due to
anomalous tunneling involving Cooper pair splitting.
\end{abstract}

\maketitle

\begin{figure}
\centering
\includegraphics[width=8cm]{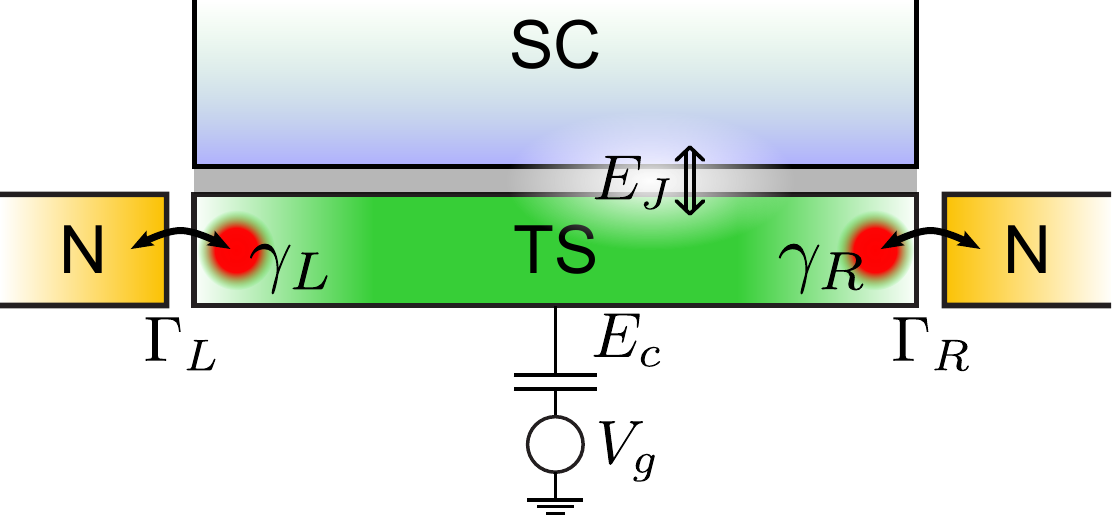}
\caption{\label{fig1} (Color online)  Majorana single-charge transistor 
(MSCT): The TS wire with Majorana end states is tunnel-coupled 
($\Gamma_L,\Gamma_R$) to normal metal electrodes and
Josephson coupled ($E_J$) to another bulk superconductor. 
Capacitive charging effects are encoded by $E_c$ and can be tuned
by a gate voltage parameter $n_g\propto V_g$. }
\end{figure}

{\it Introduction.---}Topologically nontrivial insulators and superconductors 
exhibit many remarkable  non-local features such as teleportation or
non-Abelian statistics \cite{hasan,qizhang}.  
For a one-dimensional topological superconductor (TS) wire, such effects
can be traced back to the existence of a zero-energy Majorana bound 
state (MBS) localized at each end \cite{kitaev,fukane,sato,beenakker,karsten}.  
When a grounded TS is weakly contacted by a normal metal,
the MBS is expected to produce a characteristic zero-bias anomaly peak 
in the tunnel conductance \cite{demler,nilsson,law,flensberg,wimmer}. 
Very recently, such a feature has been experimentally observed in
tunnel spectroscopy using InSb or InAs nanowires 
\cite{leo,exp1,exp2,exp3}, where Majorana fermions are theoretically expected
due to the interplay of strong 
spin-orbit coupling, Zeeman field, and proximity-induced superconducting 
pairing \cite{lutchyn,oreg,alicea}.  
Recent reviews \cite{hasan,qizhang,beenakker,karsten,alicea} have 
also summarized alternative MBS proposals.  Here we 
discuss an interacting variant of previously studied Majorana wire set-ups,
the floating ``Majorana single-charge transistor'' (MSCT)
schematically shown in Fig.~\ref{fig1}.
A comprehensive picture of its transport properties in the 
presence of interactions emerges from our analysis below.
Noting that the experimentally observed peak features could be related to a 
disorder-induced spectral peak \cite{alex,alex1},  
our results should help to distinguish the Majorana state from alternative 
explanations in future experiments.

Previous works \cite{loss1,stoud,rosch} have studied electron-electron 
interactions in an isolated TS wire and found that Majoranas still exist
under rather general conditions.  As sketched in Fig.~\ref{fig1}, we instead
study a generalization of the set-up in Ref.~\cite{leo}, where source and 
drain metallic electrodes contact the TS wire.
We stress that the MSCT could be realized not only with nanowires but 
using most other Majorana proposals as well.  In such a geometry, 
Coulomb blockade effects due to 
the finite charging energy $E_c$ of the TS  can play a decisive role. 
For instance, one expects Coulomb oscillations of the conductance 
as a function of a gate voltage parameter $n_g$, 
with peaks (valleys) near half-integer (integer) $n_g$, while
in the noninteracting ($E_c=0$) limit, the MBSs pinned to zero energy cause
resonant Andreev reflection (AR) \cite{demler,nilsson,law,flensberg},
with $n_g$-independent linear conductance $G=2e^2/h$ at temperature $T=0$.  
Resonant AR also survives for $E_c\alt \Gamma=\Gamma_L+\Gamma_R$,
albeit with reduced conductance \cite{zazu}.  
For $E_c\gg \Gamma$, Coulomb blockade is firmly established,
 and the peak conductance approaches the (spinless) resonant tunneling value 
$G=e^2/h$, which has been pointed out as
 a signature of electron teleportation \cite{fu}.  

In this paper, we consider Coulomb blockaded charge
transport through the MSCT; for a variant with one superconducting and
 one metallic lead, see Ref.~\cite{fazio}. 
We provide an exact expression for the current in this interacting system,
and develop three different approximation schemes to study
Coulomb oscillations in the MSCT both for $T=0$ and finite $T$.  
We quantitatively describe the $T=0$ crossover of the peak 
conductance from $G=2e^2/h$ to $e^2/h$ as $E_c/\Gamma$ increases, 
which constitutes a characteristic signature of Majoranas.
Remarkably, this ``halving'' of the peak
conductance is universal and found to hold for arbitrary $T$. 
For the valley conductance,  
we find that elastic cotunneling dominates while AR is subleading. 
We predict finite-voltage sidebands in the nonlinear differential conductance
which are directly related to anomalous tunneling processes where the 
Majorana state and the Cooper pair number change simultaneously. 
The presence of Majoranas can be unambiguously identified in experiments 
by the magnetic field dependence of the sideband location.

{\it Model.---}The MSCT Hamiltonian, $H=H_c+H_t+H_l$, contains a
piece $H_c$ describing the TS wire, a tunnel Hamiltonian $H_t$ 
connecting the TS to the left ($j=L$) and right ($j=R$) electrode, 
and a term $H_l$ describing the leads 
(we often use units with $e=\hbar=k_B=1$). Topological arguments warrant that
the TS wire holds a single unpaired MBS 
near each end \cite{beenakker,karsten} described by the Majorana 
operator $\gamma_j=\gamma_j^\dagger$
with anticommutator $\{\gamma_j,\gamma_{j'}\}=\delta_{jj'}$. 
We introduce the non-local fermion 
operator $d= (\gamma_L+i\gamma_R)/\sqrt{2}$ such that
$\gamma_L= (d+d^\dagger)/\sqrt{2}$ and 
$\gamma_R= -i(d-d^\dagger)/\sqrt{2}$. 
With $\hat n_d=d^\dagger d$ and the number operator $\hat N$ for
Cooper pairs in the TS,
the instantaneous charge state of the wire is described by $(N,n_d)$, 
where the integer $N$ and $n_d=0,1$ are eigenvalues of 
$\hat N$ and $\hat n_d$, respectively.  With the phase $\chi$ conjugate
to $\hat N$, where $[\chi,\hat N ] = i$ and $e^{-i\chi}$ ($e^{i\chi}$) 
lowers (raises) $N$ by one unit, we have 
\begin{equation}\label{hc}
H_c = E_c (2\hat N+ \hat n_d -n_g)^2 - E_J \cos (\chi-\phi_S).
\end{equation}
The TS wire is assumed sufficiently long to exclude a 
direct tunnel coupling between $\gamma_L$ and $\gamma_R$.  
However, note that $E_c$ introduces a dynamical coupling between the
Majoranas.  Proximity-induced pairing correlations are
required for MBS formation, and in Eq.~\eqref{hc} we include
Cooper pair exchange (with Josephson coupling $E_J$)
between the TS condensate and another bulk superconductor (with fixed 
phase $\phi_S$) \cite{fazio,zazu2}. 
We focus on the most interesting case of a large proximity gap 
$\Delta_{\rm TS}> {\rm max}(E_c,\Gamma,T)$,  where
charge transport involves MBSs and the contribution
of quasi-particles above the gap can be neglected.
Next, electrons in lead $j$ correspond to free fermions with chemical 
potential $\mu_j$ and (effectively spinless \cite{zazu}) fermionic
operators $c_{j,k}$ for momentum $k$. 
$H_l$ is treated within the usual wide-band approximation \cite{foo0}
and the bias voltage is $eV= \mu_L-\mu_R$. Taking into
account charge conservation and expressing the Majoranas in terms of
the non-local $d$ fermion, the tunnel Hamiltonian reads \cite{zazu}
\begin{equation}\label{ht1}
H_t = \sum_j \lambda_j c_{j}^\dagger \eta_j + {\rm h.c.},\quad
\eta_j= \frac{1}{\sqrt{2}}(d+s_j e^{-i\chi} d^\dagger),
\end{equation}
where $c_j=\sum_k c_{j,k}$, $s_L=+1$ and $s_R=-1$, and $\lambda_{L,R}$
denotes the respective tunnel matrix elements \cite{flensberg}.
Tunneling from the TS to lead $j$ thus proceeds either 
by destroying the $d$ state without changing $N$ (``normal'' tunneling)
or by occupying the $d$ state and simultaneously
splitting a Cooper pair (``anomalous'' tunneling), plus the
conjugate processes.  Below we use the hybridization scales
$\Gamma_j =2 \pi \nu_j |\lambda_j|^2$,
where $\nu_j$ is the density of states in lead $j$. 
Experimentally, the $\Gamma_j$ (and $n_g$) 
can be changed via gate voltages \cite{leo}.

{\it Exact expression for current.---}Using non-equilibrium 
Green's function (GF) techniques \cite{naz,alex2}, the current $I_j$ 
flowing from lead $j$ to the TS can be
expressed in terms of the Keldysh GF 
${\check G}_{\eta_j}(t,t') = -i \langle {\cal T}_C \eta_j(t) 
\eta_j(t')\rangle$,
where ${\cal T}_C$ denotes Keldysh time ordering
and the pseudo-fermion $\eta_j$ has been defined in Eq.~\eqref{ht1}.
With the Fourier-transformed retarded, 
$G_{\eta_j}^R(\epsilon)$, and Keldysh, $G^K_{\eta_j}(\epsilon)$, components
of ${\check G}_{\eta_j}$,  we obtain
$I_j = (e\Gamma_j/h) \int d\epsilon 
[ F(\epsilon-\mu_j) \ {\rm Im} G^R_{\eta_j}(\epsilon) + (i/2)
G^K_{\eta_j}(\epsilon) ]$,
where $F(\epsilon)=1-2f=\tanh(\epsilon/2T)$ 
encodes the Fermi function $f(\epsilon)$ in the leads.
Next we note that $G^K_{\eta_j}(t,t)=0$ as a consequence of
$\eta_j^\dagger \eta^{}_j=\eta^{}_j\eta_j^\dagger =1/2$.
 Hence we find the exact result 
\begin{equation} \label{current-formula}
I_j = \frac{e\Gamma_j}{h} \int d\epsilon 
F(\epsilon-\mu_j) \ {\rm Im} G^R_{\eta_j}(\epsilon),
\end{equation}
stating that the current can be computed from the spectral 
function $\propto \mbox{Im} G^R_{\eta_j}$.  The well-known expression 
for interacting quantum dots \cite{meir} has thereby been extended
to the interacting Majorana wire; note that there are two  
spectral functions associated to the currents $I_L$ and $I_R$. 
Current conservation here implies $I_L+I_R+I_S=0$, with the supercurrent
$I_S$ flowing through the interface to the bulk superconductor.
Below we define the conductance $G=d I/d V$ using the 
symmetrized current $I=(I_L-I_R)/2$.
For $E_c=0$, the pseudo-fermions $\eta_j$ reduce to Majorana fermions
$\gamma_j$, and the Lorentzian spectral function,
$-{\rm Im} G^R_{\gamma_j}(\epsilon)=\Gamma_j/(\epsilon^2+\Gamma_j^2)$,
implies resonant AR with $G=2e^2/h$ \cite{demler,nilsson,law}.
For finite $E_c$, we shall present several complementary approximations
in order to achieve a broad physical understanding of the MSCT
transport properties.  Equation \eqref{current-formula} should also 
be useful for numerically exact calculations, e.g., 
using the numerical or density-matrix renormalization group.

\textit{Equation-of-motion (EOM) approach.---}We constructed an EOM 
approach for $G^R_{\eta_j}$ to access the linear conductance near a peak. 
Within this method, we introduce the Nambu spinors
$\Psi_d = \left(d , e^{-i\chi}d^{\dagger}\right)^T$ and the
corresponding retarded GF, $G^R_{dd} = -i\Theta(t-t')
\langle \{\Psi_d(t),\Psi_d^{\dagger}(t')\}\rangle$. The EOM
for $G^R_{dd}$ generates higher-order GFs
of the type $\Gamma^R_{N^m d d} = -i\Theta(t-t')
\langle \{\hat N^m(t)\Psi_d(t),\Psi_d^{\dagger}(t')\}\rangle$, which
we truncate at the level $m=2$ and 
solve in a self-consistent way \cite{suppl}.
The resulting GF $G^R_{dd}$ then yields
$G^R_{\eta_j} = \frac12 \mbox{Tr}\left[\left(1+s_j\sigma_x\right) 
G^R_{dd}\right]$ with Pauli matrix $\sigma_x$. Finally,
we obtain the conductance from Eq.~\eqref{current-formula}. 
This approximation is valid by construction for 
$E_c \agt \Gamma$, but the imposed self-consistency \cite{suppl}
allows us to extend it to $E_c < \Gamma$, where the resulting 
conductance (being determined by trunctated fluctuations) gives a 
lower bound for the exact result.   

\textit{Zero-bandwidth model (ZBWM).---}Next we study the ZBWM 
where each lead is represented by just a single fermion site and only a 
finite number of TS Cooper pairs ($N < N_{\rm max}$) is included.  
The Hilbert space then has the finite dimension $8 N_{\rm max}$,
which allows us to numerically calculate
the spectral density $\propto \mbox{Im} G^R_{\eta_j}(\epsilon)$ via its
Lehmann representation, with poles phenomenologically broadened by $\Gamma$. 
A similar description has been pursued before for $E_c=E_J=0$ \cite{dass}.  
With this spectral function, Eq.~\eqref{current-formula} yields
the conductance within the ZBWM.

\textit{Master equation and cotunneling processes.---}For $T> \Gamma$, the GF 
formulation reduces to a master equation description including sequential 
tunneling and cotunneling processes (for simplicity, $E_J=0$ here).  
The stationary probability distribution $P_Q$ for having
 $Q=2N+n_d$ particles on the TS then obeys 
$\sum_{Q'\ne Q} [P_{Q'}W_{Q'\to Q}-P_Q W_{Q\to Q'}]=0$.
All non-vanishing transition rates $W_{Q\to Q'}$ are specified 
in terms of rates obtained under a systematic second-order 
$T$-matrix expansion in $\Gamma_{L,R}$ \cite{foot1}. 
With the electrostatic energy $E_Q=E_c(Q-n_g)^2$, 
sequential tunneling yields the rate $\Gamma^{\rm (seq)}_{j,Q\to Q \pm 1}= 
(\Gamma_j/2) f(E_{Q\pm 1}-E_Q \mp \mu_j)$ 
for one particle tunneling into (out of) the TS from (to) lead
$j=L,R$.  Next, elastic cotunneling transfers a particle from lead $j$ 
to the opposite lead $-j$ with virtual excitation of the TS states $Q\pm 1$. 
The elastic cotunneling rate is
\begin{eqnarray}\label{ec}
&& \Gamma_{j,Q}^{\rm (EC)}=\frac{\Gamma_L\Gamma_R}{8\pi}
\int d\epsilon f(\epsilon-\mu_j) [1-f(\epsilon-\mu_{-j})] 
\\ \nonumber
&&\times \left|\frac{1}{\epsilon-(E_{Q+1}-E_{Q})+i0} -
\frac{1}{\epsilon-(E_Q-E_{Q-1})-i0}\right|^2,
\end{eqnarray}
where the two terms come from the interference of normal and anomalous 
tunneling.  We note in passing that for large $\Delta_{\rm TS}$, 
inelastic cotunneling does not contribute at all, while the conventional
elastic cotunneling rate due to quasi-particle states above the gap 
(and without MBSs) would be much smaller, 
$\Gamma^{\rm (EC)}\propto \Gamma_L\Gamma_R/\Delta_{\rm TS}$ \cite{naz}.
To the same order in $\Gamma_{L,R}$, we also have local (and crossed) 
AR processes, where an electron and a hole from the same (different) 
lead(s) are combined to form a Cooper pair, $Q\to Q+2$;  
the reverse process describes Cooper pair splitting, $Q\to Q-2$.  
Some algebra yields the AR rates 
\begin{eqnarray}\label{ratear}
&& \Gamma_{j,j',Q\to Q\pm 2}^{{\rm (AR)}} =  \frac{1+\delta_{j,-j'}}{2}
\frac{\Gamma_j\Gamma_{j'}}{8\pi} \int d\epsilon \int d\epsilon' 
\\ \nonumber  && \times f(\pm(\epsilon-\mu_j)) f(\pm(\epsilon'-\mu_{j'}))
\delta(\epsilon+\epsilon'\mp (E_{Q\pm 2}-E_Q)) \\ \nonumber  && \times
\left| \frac{1}{\epsilon\mp (E_{Q\pm 1}-E_Q)+i0 }
-\frac{s_{j}s_{j'}}{\epsilon'\mp(E_{Q\pm 1}-E_Q)+i0 }\right|^2,
\end{eqnarray}
where $j=j'$ ($j\ne j'$) corresponds to local (crossed) AR. 
The $i0$ terms indicate that regularization of the integrals in 
Eqs.~\eqref{ec} and \eqref{ratear} is necessary. Application of the general 
regularization scheme in Refs.~\cite{matveev,oppen} then implies that
the principal value of these integrals needs to be taken.
Given these rates and the (numerical) solution $P_Q$ of 
the master equation, the currents $I_j$ then readily follow.

\begin{figure}
\centering
\includegraphics[width=8cm]{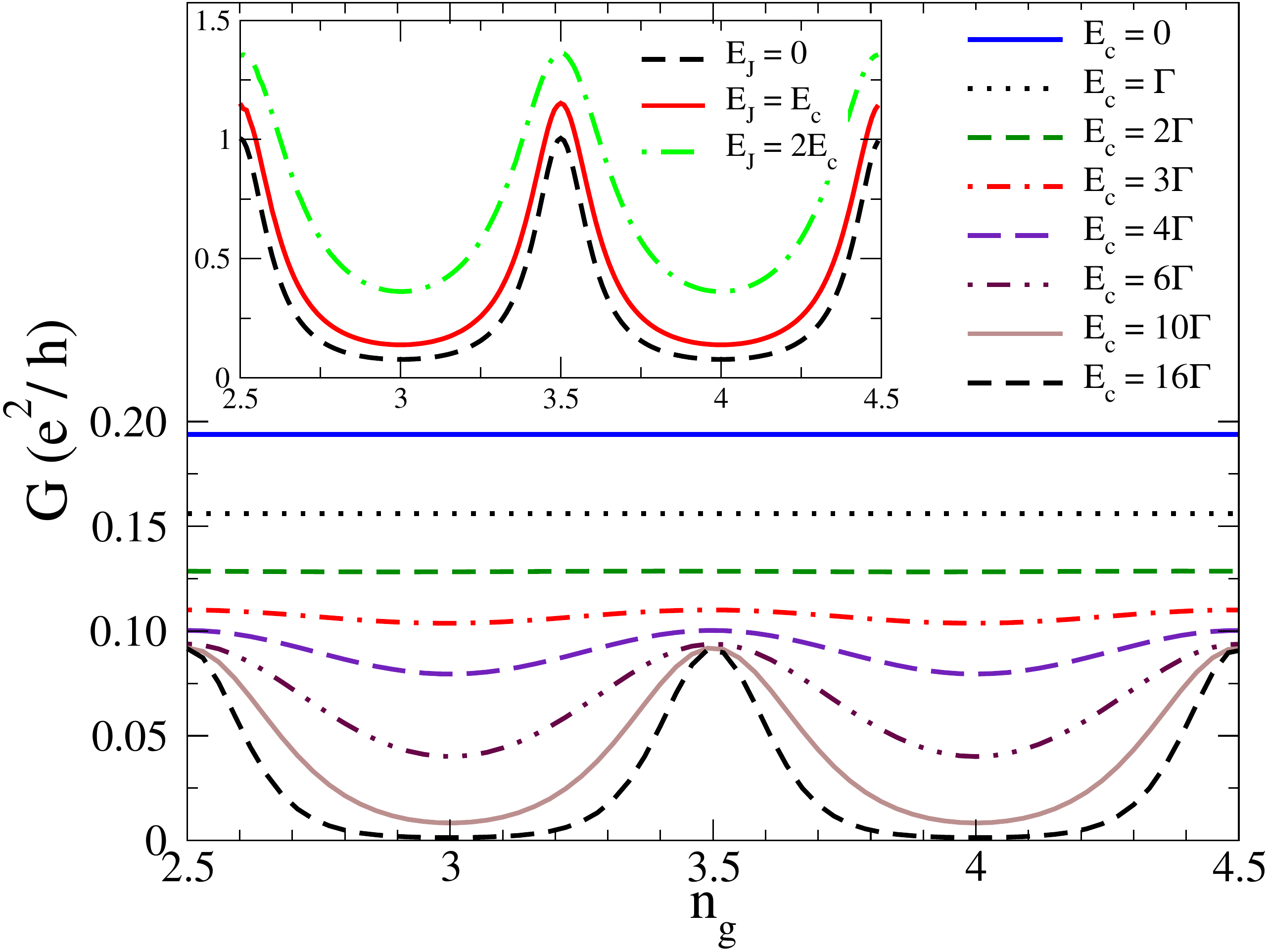}
\caption{\label{fig2} (Color online) Coulomb oscillations in the MSCT.
Main panel: Conductance $G$ vs $n_g$ from master equation for
 $E_J=0, T=2\Gamma$ and several $E_c$.  
Inset: Same but from ZBWM for $T=0, E_c=5\Gamma$ and several $E_J$. }
\end{figure}

\begin{figure}
\centering
\includegraphics[width=8cm]{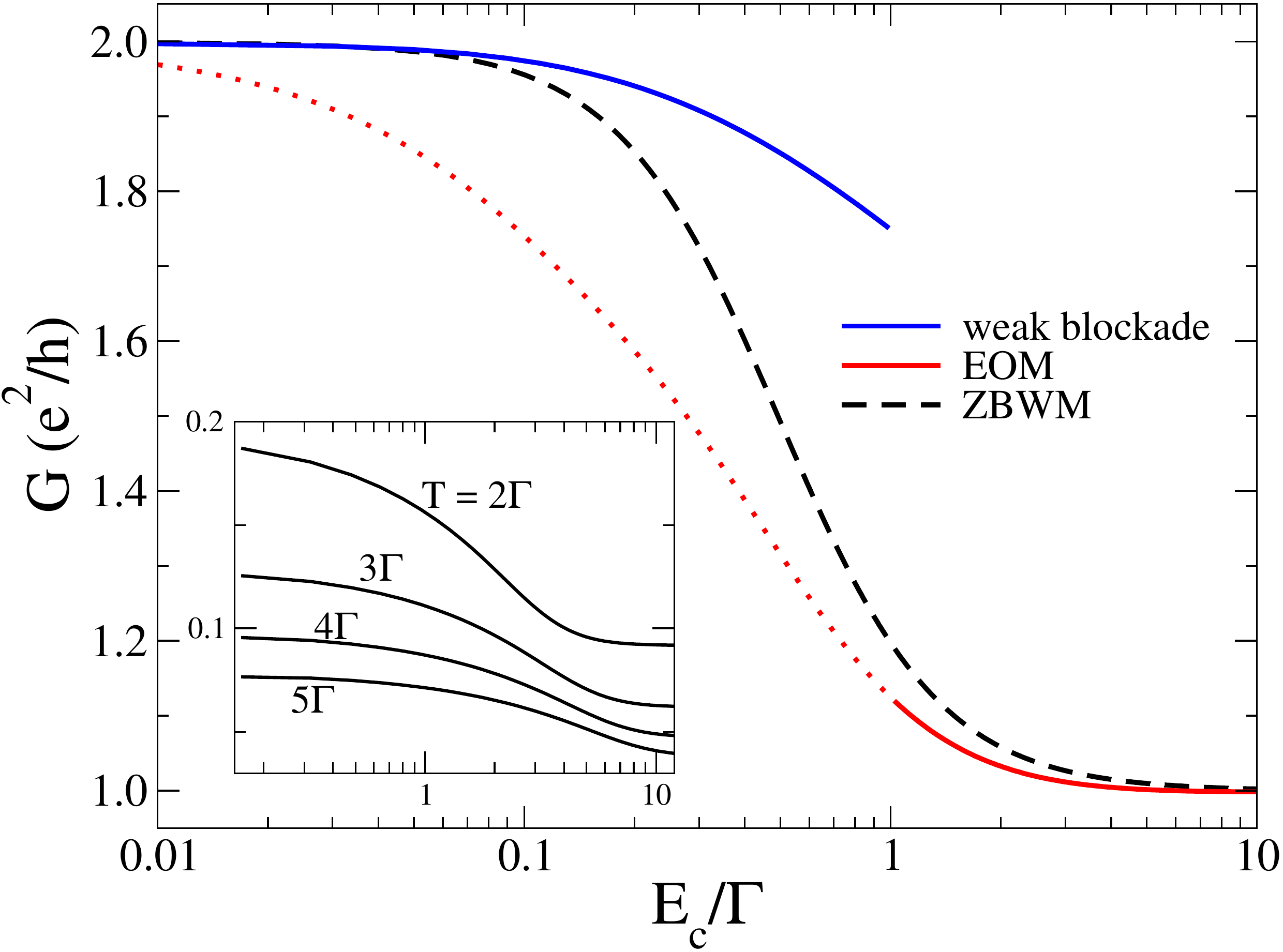}
\caption{\label{fig3} (Color online)  Peak conductance $G$ vs $E_c/\Gamma$ 
on a semi-logarithmic scale. 
Main panel: Comparison of  $T=0$ results using perturbation theory in
$E_c/\Gamma$ \cite{zazu} (blue solid curve),  the EOM approach 
(red dotted-solid curve), and the ZBWM (black dashed curve).  
The shown EOM results are quantitatively valid only for $E_c\agt \Gamma$ 
(solid part) but give a lower bound when $E_c\alt\Gamma$ (dotted part).
Here $E_J=0$ since $G$ only weakly depends on $E_J$ for $E_J\alt E_c$.  
Inset: Same but from master equation for several temperatures $T>\Gamma$.  }
\end{figure}

\textit{Coulomb oscillations.---}Let us first address the $n_g$-dependence of 
the linear ($V\to 0$) conductance, see Fig.~\ref{fig2}; we take 
$\Gamma_L=\Gamma_R=\Gamma/2$ in all figures.  Both the master equation 
(main panel, finite $T$) and the ZBWM (inset, $T=0$) reveal 
clear conductance oscillations in the MSCT for $E_c\gg \Gamma$, 
with peaks (valleys) for half-integer (integer) gate voltage parameter $n_g$.  
The main panel shows that the peak (valley) conductance is halved 
(strongly suppressed) when going
from the noninteracting to the deep Coulomb blockade limit. 
For $E_J=0$ and $(\Gamma, T)\ll E_c$, the lineshape of 
the valley conductance is obtained in closed form,
\begin{equation}\label{valley}
G_{\rm valley}(\delta) = \frac{e^2}{h} \frac{\Gamma_L\Gamma_R}{E_c^2}
\frac{1}{(1-4\delta^2)^2},
\end{equation}
where $\delta=n_g-[n_g]$ with $|\delta|\ll 1$ is the deviation from
a valley center.
Equation \eqref{valley} comes from elastic cotunneling,
with constructive interference of the
normal and anomalous tunneling contributions [see Eq.~\eqref{ec}], 
while AR is strongly suppressed in this limit.
The inset of Fig.~\ref{fig2} shows that $G$ increases when 
the Josephson coupling $E_J$ grows. One can understand this by 
noting that for $E_J\gg E_c$, one ultimately reaches the resonant 
AR limit of a grounded TS, with the $n_g$-independent $T=0$ 
conductance $G=2e^2/h$. We find that AR yields significant 
conductance contributions for $E_J\agt E_c$, which
are best detected through the non-local conductance
$\partial I_L/\partial \mu_R$. However, we will discuss this
quantity in detail elsewhere.

\textit{Peak conductance.---}Results for the peak conductance
are shown in Fig.~\ref{fig3}.  For $T=0$ (main panel), we obtain the full
crossover from $G=2e^2/h$ to $G=e^2/h$ as $E_c/\Gamma$ increases.  
The known small-$E_c$ behavior \cite{zazu} is nicely reproduced by the 
ZBWM calculation. In the opposite large-$E_c$ limit, 
the EOM method is very accurate and Fig.~\ref{fig3} suggests that 
the  simple ZBWM already captures the crossover 
from resonant AR \cite{demler,nilsson,law} to electron teleportation \cite{fu}
surprisingly well.   The inset of Fig.~\ref{fig3} again demonstrates
the universal halving of the finite-$T$ peak conductance, 
see also Fig.~\ref{fig2}.  Since experiments so far were 
conducted in the high-temperature regime $T>\Gamma$ \cite{leo}, 
let us now specify the lineshape near a conductance peak for $E_c\gg \Gamma$.
Using $\delta=n_g-[n_g]-1/2$ with $|\delta|\ll 1$ for the deviation
from a peak center, truncation of the master equation to two charge states 
gives
\begin{equation}\label{pcon}
G_{\rm peak}(\delta) =\frac{e^2}{h} \frac{\pi\Gamma}{16T} \frac{1}
{\cosh^2(\delta E_c/T)}. 
\end{equation} 
We stress that the peak conductance [$G_{\rm peak}(\delta=0)$] 
is indeed halved compared to $E_c=0$ \cite{demler}. Moreover,
it exhibits both a thermal and an interaction-induced reduction.

\begin{figure}
\centering
\includegraphics[width=8cm]{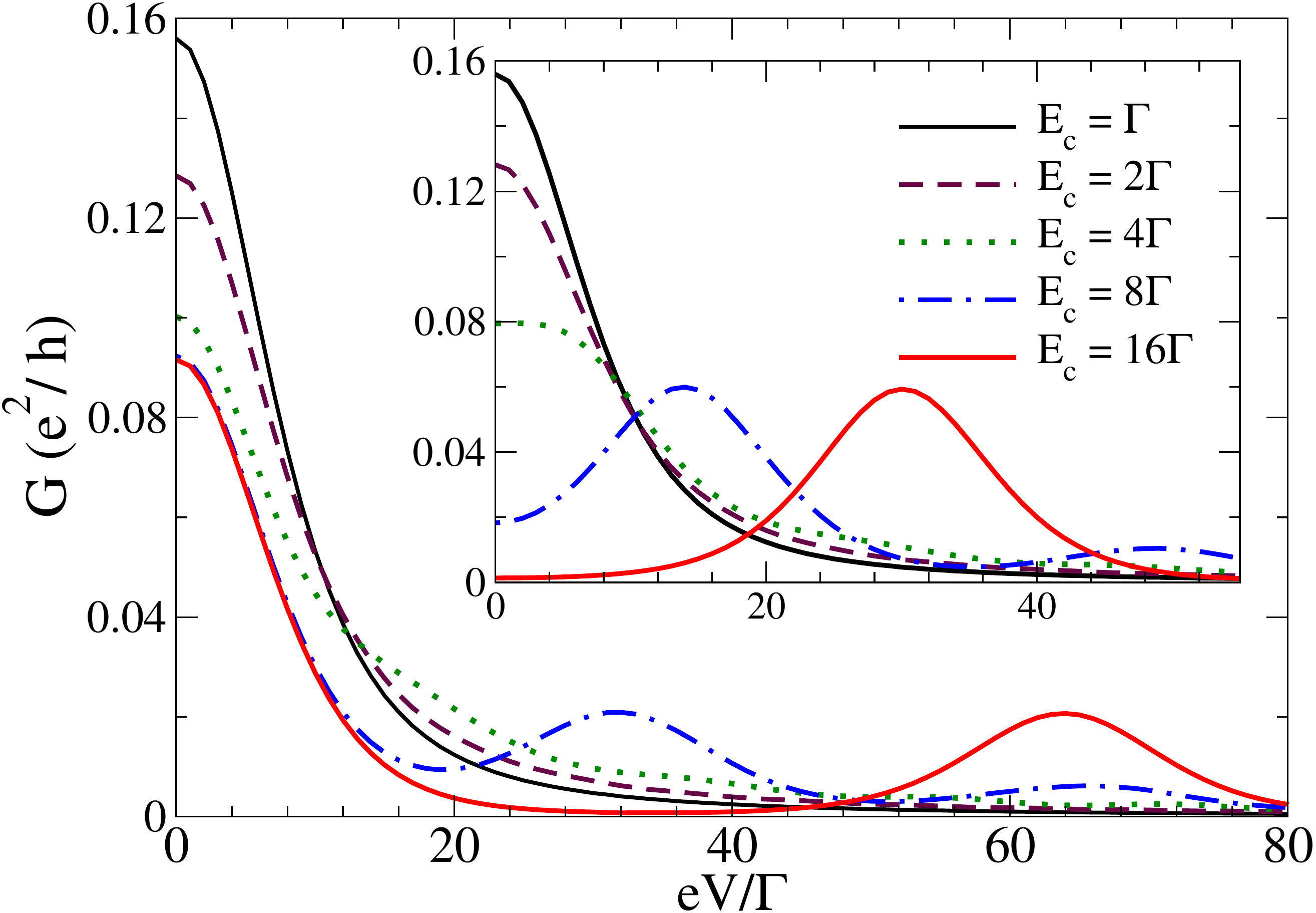}
\caption{\label{fig4} (Color online) 
$G=dI/dV$ vs voltage $V$ from the master equation for $T=2\Gamma$,
$E_J=0$, and several $E_c/\Gamma$.
The main panel (inset) is for half-integer (integer) $n_g$.
 }
\end{figure}

\begin{figure}
\centering
\includegraphics[width=8cm]{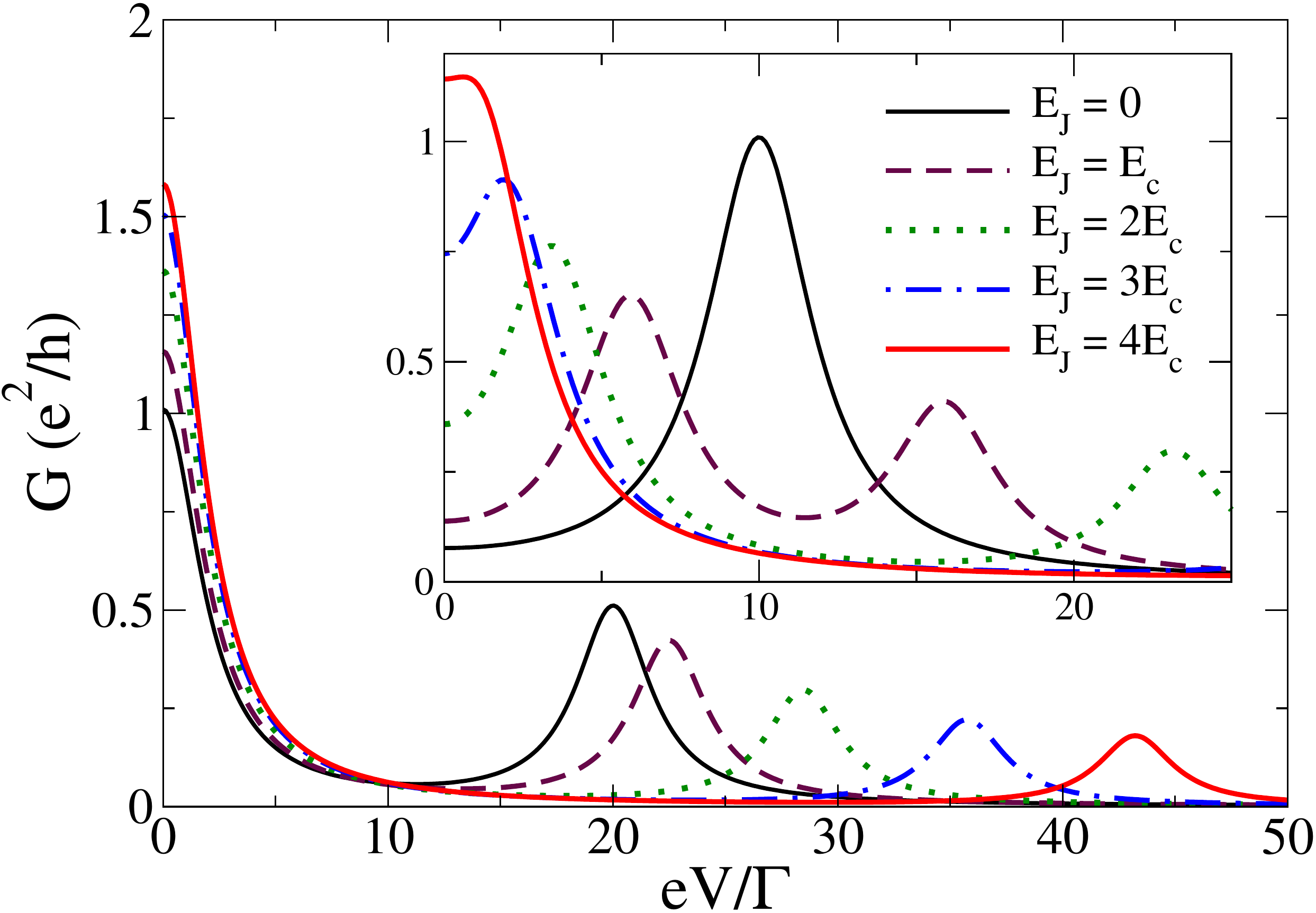}
\caption{\label{fig5} (Color online)  Same as Fig.~\ref{fig4} but
from ZBWM for $T=0$, $E_c=5\Gamma$ and several $E_J$. 
}
\end{figure}

\textit{Finite-voltage sidebands.---}Next we discuss the
differential conductance at finite bias voltage $V$.  
Master equation results for $T=2\Gamma$ are shown in Fig.~\ref{fig4}.
Starting with the main panel, we find sideband peaks
when $eV$ is equal to an integer multiple of $4E_c$.
For these voltages, the chemical potentials $\mu_{L,R}$ are resonant
with two (almost) degenerate higher-order charge states, 
implying additional sequential tunneling contributions
beyond the resonant transition determining the linear conductance peak
[Eq.~(\ref{pcon})].  Note that the fluctuations in $N$ needed to reach 
higher-order charge states can only be achieved through anomalous
tunneling processes [see Eq.~\eqref{ht1}]. Similar sideband peaks are also
found for other $n_g$; the integer-$n_g$ valley case is shown in the inset 
of Fig.~\ref{fig4}.  In Fig.~\ref{fig5} we show the evolution of the 
sideband peaks as $E_J$ is changed, determined from the ZBWM at $T=0$.
For half-integer $n_g$, the sideband peak position
observed in the main panel of Fig.~\ref{fig5} 
is well described by $eV\simeq  4E_c\sqrt{1+(E_J/2E_c)^2}$,
which comes from Josephson coupling between the two relevant charge states.
Since $E_J$ can be tuned by applying a
small magnetic field parallel to the junction between the TS and the
bulk superconductor, an experimental observation of the sideband peak 
and its shift with magnetic field [cf.~the expression for the peak position above] would provide clear
evidence for the anomalous tunneling mechanism, and thereby for the
presence of MBSs.  

\textit{Conclusions.---}In this paper, we have studied the
transport properties of an interacting Majorana single-charge transistor.
Our results should be directly relevant for experiments 
extending existing work, see, e.g., Ref.~\cite{leo}, where 
conductance peaks for tunneling into Majorana wires were reported. 
When a gate voltage parameter $n_g$ is varied, we
find Coulomb oscillations, where the behavior of the peak
and valley conductance has been characterized in detail.
The Majorana fermions in this system could be identified by observing 
sideband peaks in the nonlinear conductance and from the crossover behavior
of the Coulomb peak conductance.---  
This work was supported by the DFG 
(Grant No.~EG-96/9-1 and SFB TR 12), by the EU network SE2ND,
and by the Spanish MICINN under contract FIS2008-04209.
\textit{Note added:} After submission of this work, we learned
of an unpublished study of the MSCT model by L. Fu and C.L. Kane.

\appendix
\section*{Supplementary material: EOM approach}

Using a notation as in the main text, the current follows from
the retarded GF
$G^R_{dd}(t) = -i \Theta(t) \mean{ \{ \Psi_d(t), \Psi_d^\dagger(0) \} },$
which is a $2\times 2$ matrix in Nambu space with
spinors $\Psi_d = (d , e^{-i\chi} d^\dagger)^T$.
In energy ($\epsilon$) space, it obeys the EOM
\begin{equation} \label{suppl:eom}
(\epsilon-E_0 + V_g + i\hat \Gamma) G^R_{dd} = 1 + 4 E_c \Gamma^R_{Ndd},
\end{equation}
where $V_g = 2E_c n_g$, $E_0 = \begin{pmatrix} E_c &0 \\ 0&3E_c\end{pmatrix}$,
and $\hat\Gamma = \sum_j \Gamma_j (1+s_j \sigma_x)$.
$\Gamma^R_{Ndd}$ in Eq. \eqref{suppl:eom} is the first in a hierarchy of vertex
functions
($m = 1, 2, \dots$),
\begin{equation}
\Gamma^R_{N^mdd}(t) = -i \Theta(t)
\mean{\{\hat{N}^m(t)\Psi_d(t),\Psi_d^\dagger(0)\}},
\end{equation}
which are generated sequentially through their EOM.
In particular,
\begin{equation} \label{suppl:eom_Gamma}
        (\epsilon-E_0 + V_g + i \hat\Gamma) \Gamma^R_{Ndd}
        = A - i \tilde{\Gamma} G^R_{dd} + 4 E_c \Gamma^R_{N^2dd},
\end{equation}
with $\tilde{\Gamma} = \sum_j \Gamma_j s_j \sigma_x$ and
\begin{equation} \label{suppl:A}
        A = \mean{\{\hat{N}\Psi_d,\Psi_d^\dagger\}}
        = \begin{pmatrix} \mean{\hat{N}} & 0 \\ 0 &
\mean{\hat{N}-(1-\hat{n}_d)} \end{pmatrix}.
\end{equation}
Next we discuss an approximation closing the above set of equations.
We see from Eqs.~(\ref{suppl:eom}) and (\ref{suppl:eom_Gamma}) that
the energy dependence of each higher-order vertex function
produces a pole in $G^R_{dd}$, and the scale of the energy spacing
between the poles is set by the prefactor $4E_c$ with which the
vertex functions appear in the EOMs.
For $E_c \gtrsim \Gamma$, the physics is therefore
controlled by a small number of poles only, which allows us to close the
EOM hierarchy by truncation (effectively keeping just a few poles).
An approximation keeping only two poles can be achieved
by imposing the variational condition
\begin{equation} \label{suppl:decoupling}
        \Gamma^R_{N^2 dd} = B \Gamma^R_{Ndd},
\end{equation}
with a Nambu matrix $B$. {}From this we obtain the closed equation
\begin{align}
        &\Bigl[ (\epsilon-E_0 + i \hat\Gamma - 4 E_c B  + V_g)
        (\epsilon-E_0 + i \hat\Gamma + V_g) \nonumber\\
        &+ 4 i E_c \tilde{\Gamma} \Bigr] G^R_{dd}
        = \epsilon-E_0 + i \hat\Gamma + 4 E_c(A- B) + V_g.
\label{suppl:eom_closed}
\end{align}
For symmetric contacts, $\Gamma_L = \Gamma_R$, we have $\tilde{\Gamma} = 0$,
$\hat \Gamma=\Gamma$, all matrices become diagonal, and
\begin{align} \label{suppl:G_dd_approx}
        G^R_{dd} = \frac{B^{-1}A}{\epsilon-E_0 + i\Gamma- 4E_c B + V_g}
        + \frac{1-B^{-1}A}{\epsilon-E_0 + i\Gamma + V_g}.
\end{align}
The vertex function correspondingly reads
\begin{equation} \label{suppl:Gamma_approx}
        \Gamma^R_{Ndd} = \frac{A}{\epsilon-E_0 + i\Gamma- 4E_c B + V_g}.
\end{equation}
This EOM implementation is valid for
gate voltages $V_g$ close to resonance for either the $[G^R_{dd}]_{11}$ or
$[G^R_{dd}]_{22}$ matrix entries.
The on-resonance entry will then be dominated by a single
central pole, while the other entry is off-resonant due to the
$2E_c$ energy shift between the two entries in the matrix $E_0$.
Through this shift, the Fermi surface lies almost in the center between
the two poles of $G^R_{dd}$, and keeping just these two poles is
sufficient for $E_c \gtrsim \Gamma$.
For $E_c < \Gamma$, however, higher-order poles become important. These
come from higher-order fluctuations of $\hat{N}^m$ in
the neglected vertex functions. Hence Eq.~(\ref{suppl:G_dd_approx})
can be interpreted as a truncation of fluctuations in the number of
Cooper pairs.  Therefore, while we obtain quantitatively accurate
conductance results for  $E_c > \Gamma$,
due to the truncation of Cooper pair fluctuations
we get only a  lower bound for the conductance when $E_c < \Gamma$.
Within the restrictions imposed by the truncation, however, we
achieve an optimal solution for $G^R_{dd}$ by
exact fulfillment of the following sum rules.

{}From Eqs.~(\ref{suppl:G_dd_approx}) and (\ref{suppl:Gamma_approx}),
a self-consistent computation of four parameters is required,
namely  $\mean{\hat{N}}$ and $\mean{\hat{n}_d}$ appearing in the matrix
$A$ [see Eq. (\ref{suppl:A})], and the two diagonal matrix entries
$B_{11}$ and $B_{22}$ of $B$ [see Eq.~(\ref{suppl:decoupling})].
(Alternatively, it can be advantageous to fix $\mean{\hat{n}_d}$ but
adjust $V_g$ self-consistently.)
These values are determined from exact sum rules,
\begin{align}
        -\frac{1}{\pi} \int d\epsilon f_d(\epsilon) \text{Im}[G^R_{dd}(\epsilon)]_{11} &= \mean{\hat{n}_d},
\\
        -\frac{1}{\pi} \int d\epsilon f_d(\epsilon) \text{Im}[G^R_{dd}(\epsilon)]_{22} &= 1-\mean{\hat{n}_d},
\\
        -\frac{1}{\pi} \int d\epsilon f_d(\epsilon)
\text{Tr}\{\text{Im}\Gamma^R_{Ndd}(\epsilon)\} &= \mean{\hat{N}-(1-\hat{n}_d)},
\end{align}
where $f_d$
is the distribution function on the TS wire; in equilibrium, $f_d=f$.
(Note that through the trace over $\text{Im}\Gamma^R_{Ndd}$,
averages of the form $\mean{\hat{N} \hat{n}_d}$ cancel out.)
However, when using Eq.~\eqref{suppl:G_dd_approx},
we have to self-consistently adjust four parameters
with three sum rules only. We thus impose
$\text{Tr}B = B_{11} + B_{22} = \mean{\hat{N}} +  (1-\mean{\hat{n}_d})$,
which reproduces in the large-$E_c$
limit the resonances found with the ZBWM and master equation approaches
around $\mean{\hat{n}_d}=1/2$.
The EOM results shown in Fig.~\ref{fig3} then follow (with $T=0$ and
given ratio $E_c/\Gamma$) by self-consistently solving for $G_{dd}^R$.

\begin{thebibliography}{99}
\bibitem{foo0}
It is straightforward to go beyond this approximation by 
allowing for energy-dependent $\Gamma_j(\epsilon)$ in the equations below.
\bibitem{foot1}
We find $W_{Q\to Q\pm 1}=\sum_j \Gamma_{j,Q\to Q\pm 1}^{\rm (seq)}$
and $W_{Q\to Q\pm 2}= \Gamma_{LR,Q\to Q\pm 2}^{\rm (AR)}+
\sum_j \Gamma_{jj,Q\to Q\pm 2}^{\rm (AR)}$.  Additional rates
(not specified here) involving Cooper pair transfer between the
TS and the bulk superconductor are included for finite $E_J$.
\end{thebibliography}
\end{document}